\documentclass[%
 reprint,
superscriptaddress,
 nofootinbib,
 amsmath,amssymb,
 aps,
 prd,
]{revtex4-2}

\usepackage{xcolor}
\usepackage{multirow}
\usepackage{tabularx}
\usepackage{graphicx}
\usepackage{dcolumn}
\usepackage{bm}
\usepackage{acronym}
\usepackage{hyperref}
\usepackage{threeparttable}
\usepackage{subcaption}

\definecolor{dodgerblue}{HTML}{1E90FF}
\definecolor{garibaldired}{HTML}{DD0000}
\definecolor{seagreen}{HTML}{00916E}

\newcommand{\portsmouth}{Institute of Cosmology \& Gravitation, University of Portsmouth, Portsmouth, United Kingdom}
\newcommand{\birmingham}{Institute for Gravitational Wave Astronomy \& School of Physics and Astronomy, University of Birmingham, Birmingham, United Kingdom}
\newcommand{\southampton}{Mathematical Sciences, University of Southampton, Southampton, United Kingdom}
\newcommand{\glasgow}{SUPA, University of Glasgow, Glasgow, United Kingdom}

\begin{document}

\preprint{APS/123-QED}

\title{Premerger observation and characterization of massive black hole binaries}

\author{Gareth {Cabourn Davies}}
\email{gareth.cabourn-davies@port.ac.uk}
\affiliation{\portsmouth}
\author{Ian Harry}%
\email{ian.harry@port.ac.uk}
\affiliation{\portsmouth}
\author{Michael J. Williams}%
\email{michael.williams3@port.ac.uk \\ These 3 authors contributed equally.}
\affiliation{\portsmouth}

\author{Diganta Bandopadhyay}
\affiliation{\birmingham}
\author{Leor Barack}
\affiliation{\southampton}
\author{Jean-Baptiste Bayle}
\affiliation{\glasgow}
\author{Charlie Hoy}
\affiliation{\portsmouth}
\author{Antoine Klein}
\affiliation{\birmingham}
\author{Hannah Middleton}
\affiliation{\birmingham}
\author{Christopher J. Moore}
\affiliation{\birmingham}
\author{Laura Nuttall}
\affiliation{\portsmouth}
\author{Geraint Pratten}
\affiliation{\birmingham}
\author{Alberto Vecchio}
\affiliation{\birmingham}
\author{Graham Woan}
\affiliation{\glasgow}

\begin{abstract}
We demonstrate an end-to-end technique for observing and characterizing massive black hole binary signals
before they merge with the LISA space-based gravitational-wave observatory.
Our method uses a zero-latency whitening filter, originally designed for rapidly
observing compact binary mergers in ground-based observatories, to be able to observe signals with no additional latency due to filter length.
We show that with minimal computational cost, we are able to reliably 
observe signals as early as 14 days premerger as long as the signal has
accrued a signal-to-noise ratio of at least 8 in the LISA data.
We also demonstrate that this method can be used to characterize the source properties, providing early estimates of the source's merger time, chirp mass, and sky localization.
Early observation and characterization of massive black holes is
crucial to enable the possibility of rapid multimessenger observations, and to ensure that LISA can enter a protected operating period when the merger signal arrives.

\end{abstract}

\maketitle

\acrodef{LISA}[LISA]{Laser Interferometer Space Antenna}
\acrodef{MBHB}[MBHB]{Massive Black Hole Binary}
\acrodef{PSD}[PSD]{Power Spectral Density}
\acrodef{ASD}[ASD]{Amplitude Spectral Density}
\acrodef{FAR}{false alarm rate}

\section{Introduction}
\label{sec:intro}

The late 2030s will see the first data from the \ac{LISA} gravitational-wave observatory~\cite{LISA:2017pwj, Colpi:2024xhw}.
\ac{LISA} will enable us to observe the gravitational-wave spectrum between $\sim 10^{-4}$~Hz and  $\sim 10^{-1}$~Hz and allow the observation of a wide range of astrophysical phenomena~\cite{LISA:2022yao}.
One of the primary science motivators for \ac{LISA}, and the focus of this work, is the observation of \ac{MBHB} mergers.
LISA will have peak sensitivity to \ac{MBHB} mergers with total mass between $10^4$ and $10^7 M_{\odot}$~\cite{LISA:2022yao, Merritt:2004gc, Kormendy:2013dxa, Colpi:2014poa, Yang:2021imt}.

The problem of observing and characterizing \ac{MBHB} mergers with \ac{LISA} has been explored over the last two decades, primarily in the context of the ongoing LISA Data Challenges (formerly called the Mock LISA Data Challenges)~\cite{Arnaud:2006gm, MockLISADataChallengeTaskForce:2006sgi, Arnaud:2007vr, MockLISADataChallengeTaskForce:2007iof, Babak:2008aa, Arnaud:2007jy, MockLISADataChallengeTaskForce:2009wir, LDC_WEBSITE}. 
Such signals are expected to have a very large signal-to-noise ratio, rendering the identification of sources trivial~\cite{LISA:2022yao}.
However, this large signal-to-noise ratio makes the characteriziation of these sources, as performed using numerical Bayesian analysis toolkits, challenging~\cite{LISA:2022yao, Arnaud:2006gm, MockLISADataChallengeTaskForce:2006sgi, Arnaud:2007vr, MockLISADataChallengeTaskForce:2007iof, Babak:2008aa, Arnaud:2007jy, MockLISADataChallengeTaskForce:2009wir, LDC_WEBSITE}.
The main problems that need to be solved in advance of LISA's launch are how to characterize these signals in the presence of a large number of other overlapping signals---\ac{LISA}'s ``Global Fit'' problem~\cite{Cornish:2005qw}---and how to produce waveforms with sufficient accuracy~\cite{Purrer:2019jcp}.

Tackling these well known problems is not the focus of this work.
The large signal-to-noise ratio of \ac{MBHB} mergers offers another interesting science possibility.
In particular, these systems may be observable in the days (or even weeks) before they merge~\cite{Lops:2022ooj,Piro:2022zos}.
If it were possible to identify these systems before they merge, and constrain the sky location where the signal originates from, it would open up the possibility of alerting electromagnetic astronomers that a \ac{MBHB} merger is about to occur~\cite{Lops:2022ooj,Piro:2022zos}.
As \ac{MBHB} mergers can occur in gas-rich environments, the gravitational-wave signal may be accompanied by EM emission~\cite{Bogdanovic:2021aav}. One of the Science Objectives for the LISA Mission is to ``Identify the electromagnetic counterparts of massive Black Hole
binary coalescences'' with a stated aim to alert astronomers hours to days before merger~\cite{Colpi:2024xhw}.
In addition, it would be very useful for operation of the \ac{LISA}
satellite to know that an \ac{MBHB} merger is imminent.
LISA science data will contain schedulded interruptions due to antenna re-pointing or discharge of test masses. 
If we know the merger time of a \ac{MBHB} in advance, LISA can enter a protected period in which scheduled interruptions are rearranged~\cite{Colpi:2024xhw}.  
Depending on ground station availability, LISA can also enter a low-latency period in which data is downloaded in quasi-real time for the merger itself~\cite{Colpi:2024xhw}.

The topic of premerger observation of \acp{MBHB} and in particular
the ability to localize \acp{MBHB} premerger has been explored 
already in the literature (for e.g.~\cite{Lang:2007ge, Kocsis:2007yu, Trias:2007fp, McWilliams:2011zs, DalCanton:2019wsr, Mangiagli:2020rwz, Lops:2022ooj,Piro:2022zos, Saini:2022hrs, Chen:2023qga,Houba:2024mqj}). However, the majority of these works
are carried out using Fisher Matrix analyses, or Bayesian inference
using waveforms that are abruptly terminated in the frequency domain.
As we will discuss and explain in detail below, traditional likelihood computations would require waiting for additional data (up to a day) to arrive before one can accurately whiten the waveforms and compute a matched filter. We do
note the recent works of~\cite{Houba:2024mqj,Ruan:2024qch}, which propose a direct
search using convolutional neural networks and could provide an alternative to the methods presented here.

In this work, we present an end-to-end method for identifying and characterizing \ac{MBHB} signals before merger.
This method incorporates methodology developed
by the GstLAL search team in the context of ground-based searches~\cite{Tsukada:2017cuf}, to allow us to analyse LISA data immediately as it arrives.
We illustrate the utility of the method by demonstrating the recovery of signals added to simulated LISA noise between 0.5 and 14 days
before merger. We also demonstrate that this method can be used to characterize these simulated signals without needing to wait for additional data to arrive, by interfacing with one~\cite{Weaving:2023fji} of the Bayesian inference toolkits now available for LISA parameter estimation~\cite{Marsat:2020rtl,Katz:2020hku,Buscicchio:2021dph,Littenberg:2023xpl, Hoy:2023ndx}.
We highlight the importance of having LISA data available below $10^{-4}\,$Hz: Many signals
will have dominant mode emission $< 10^{-4}\,$Hz in the days before merger, and so having sensitivity below this will be crucial for
early identification and localization.

The layout of this work is as follows:
In section \ref{sec:motivation} we explore the motivation for this work and describe the data and noise curves that we will consider.
In section \ref{sec:search_method} we describe our new technique for observing and characterizing \ac{MBHB} systems in the days to weeks before merger, applying techniques developed in the LIGO-Virgo-KAGRA (LVK)  context to achieve minimal latency.
In section \ref{sec:search_results} we demonstrate that we can effectively observe \ac{MBHB} mergers, as soon as the accrued signal-to-noise ratio becomes larger than $8$, with insignificant computational cost.
In section \ref{sec:PE} we apply Bayesian parameter inference tools to demonstrate that this method can be used to provide rapid premerger constraints on the value of the MBHB's sky location, chirp mass and time of arrival.
Finally, we conclude in section \ref{sec:conclusion}. 
All of our results and figures are fully reproducible, with the code
used to make them publicly available. Please see our \url{https://icg-gravwaves.github.io/lisa_premerger_paper/} data release page for more detail.

\section{Motivation and setup}\label{sec:motivation}

In this section we will demonstrate the potential of premerger observation of \ac{MBHB} mergers with \ac{LISA}, and describe the assumptions on the LISA sensitivity that will be used throughout this work.
Other works have previously noted the potential for observing \ac{MBHB} systems before merger (e.g.,~\cite{Lang:2007ge, Kocsis:2007yu, Trias:2007fp, McWilliams:2011zs, Mangiagli:2020rwz, Lops:2022ooj,Piro:2022zos,Saini:2022hrs, Chen:2023qga,Houba:2024mqj}), but we independently repeat and review that motivation in this section for completeness.
In later sections we will present our practical novel method for performing such observations.

\subsection{Sensitivity curve}

\begin{figure}
    \includegraphics{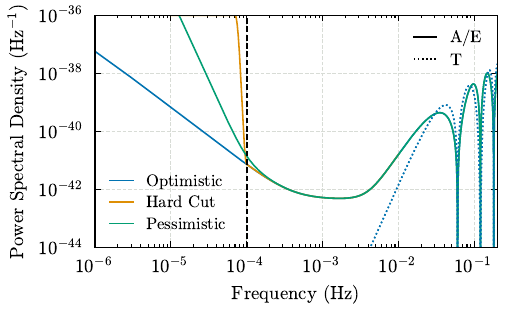}
    \includegraphics{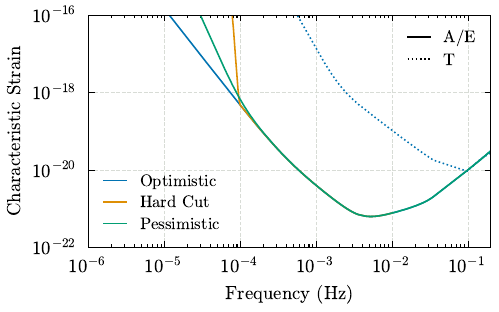}
    \caption{The LISA sensitivity curves for the A (and equivalently E) TDI channels used in this work. Top: The power spectral density of the channels. Bottom: The characteristic strain. The optimistic curve smoothly continues the low-frequency sensitivity below $1\times10^{-4}$Hz. The pessimistic curve includes a restricted sensitivity below $1\times10^{-4}$Hz and the hard cut allows for no sensitivity below $1\times10^{-4}$Hz. We remind the reader that the A,E and T channels do not directly correspond to gravitational-wave strain and there is a response function to translate from one to the other. Although the PSD of the T channel goes to 0 at low frequencies, it is insensitive to gravitational waves there as shown in the lower panel. Note that we use ``TDI 1.5'' combinations in this work~\cite{Otto:2015erp}.}
  \label{fig:psd}
\end{figure}

Before we can motivate the potential of an early-warning search
for \ac{MBHB} mergers, we must define the noise performance of the
LISA instrument.  LISA's definition study report~\cite{Colpi:2024xhw}
illustrates LISA's predicted sensitivity curve, but highlights that
the performance below $10^{-4}$Hz is an extrapolation. Much of LISA's science
objectives does not depend on the sensitivity below $10^{-4}$Hz, but, as
we will show, premerger observation of \ac{MBHB} systems is sensitive to
this. The dominant emission mode for \ac{MBHB} mergers will often be at less than $10^{-4}$Hz in the 30 days preceding merger.

Therefore, in this work we consider three potential sensitivity curves to explore the potential effect of including, or not, data below 
$10^{-4}\,$Hz. We use an ``optimistic'' curve where, following Figure 2.2
of~\cite{Colpi:2024xhw}, we smoothly extend the slope of the noise curve at low frequencies.
We also consider a ``pessimistic'' noise curve, where there is sensitivity below $10^{-4}\,$Hz, but the slope of the \ac{PSD} is increased with respect to the ``optimistic'' noise curve.
Finally, we  consider a ``hard cut'' PSD, where there is no sensitivity below $1\times10^{-4}\,$Hz\footnote{To avoid adding a Heaviside step discontinuity in the PSD we implement this as a rapidly increasing sigmoid function in the PSD below $10^{-4}$Hz.}.
These three noise curves, quantified both in terms of the \ac{PSD} in the A, E, T time-domain interferometry (TDI) channels, and the characteristic strain in these channels, are shown in Figure~\ref{fig:psd}.
We note that we consider noise
in the A, E and T channels to be independent and we use ``TDI 1.5'' throughout this work, as described in~\cite{Otto:2015erp}.

\begin{figure}
    \includegraphics{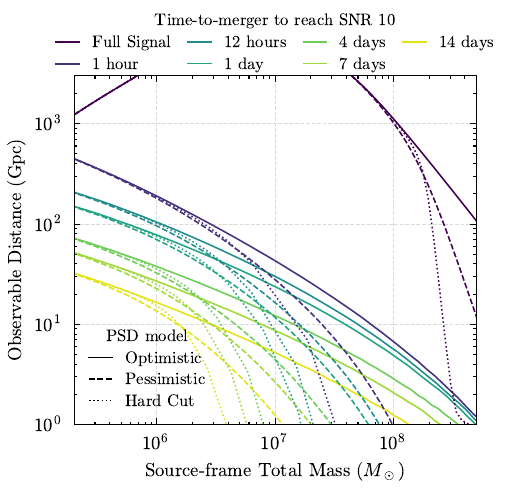}
    \includegraphics{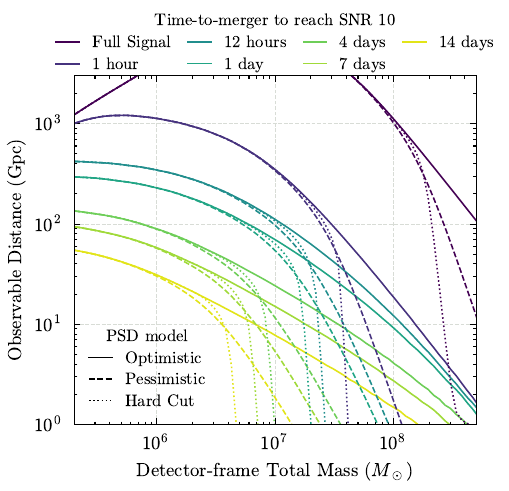}
    \caption{Observable distance for equal mass systems as a function of total mass for different premerger times at a signal-to-noise ratio of 10. Top: detector-frame masses Bottom: source frame masses.}
  \label{fig:motivation}
\end{figure}

\subsection{Ability to observe MBHB systems before merger}

We now explore the potential of observing \ac{MBHB} systems
in the days before merger. To do this we need to compute the distance
to which we can observe \ac{MBHB} systems as a function of time before
merger, assuming we are able to recover all of the signal power accumulated up until that point.

To compute this we make use of an inner product that is standard in gravitational-wave astronomy, the matched filter,
\begin{equation}
\label{eq:mf_standard}
    (a | b) = 4 \int_{0}^{\infty} \frac{\tilde{a}^{\star}(f) \tilde{b}(f)}{S_\textrm{n}(f)} \textrm{d}f,
\end{equation}
where $\tilde{a}$ and $\tilde{b}$ represent the Fourier transform of two timeseries $a$ and $b$, $S_n(f)$ represents the \ac{PSD} and $\tilde{a}^{\star}$ represents a complex conjugate of $\tilde{a}$.

The optimal signal-to-noise ratio ($\rho$) of a signal observed in LISA is then given by
\begin{equation}
\rho = \sqrt{ (h_{A}| h_{A}) + (h_{E}| h_{E}) + (h_{T}| h_{T})}.
\end{equation}
Here, $h_{X}$ is the gravitational-wave signal, as it would appear in the TDI channels (A, E or T). 
For this section, we consider a \ac{MBHB} system ``observable'' if it has an optimal signal-to-noise ratio of 10 or larger\footnote{We will properly compute the detection threshold of premerger MBHB signals later in this work. 10 will suffice for our general motivation at this stage.}.

The optimal signal-to-noise ratio of a signal will depend on the 15 parameters that are required to fully describe a \ac{MBHB} system on a circular orbit.
For simplicity, we will neglect the component spins and assume the system has equal mass. We can then compute an average ``observable distance'' $d_{\rm obs}$ as a function of the total mass $M$. This is done by simulating a set of 200 systems with the orientation and sky location drawn from an isotropic distribution, computing the signal-to-noise ratio at a reference distance, averaging over the 200 systems, and converting the averaged signal-to-noise ratio into a distance at which it would be 10.
The optimal signal-to-noise ratio is computed by generating waveforms using the \texttt{BBHx} package~\cite{Katz:2020hku, Katz:2021uax, michael_katz_2023_7791640}, which produces waveforms from the PhenomD waveform family~\cite{Khan:2015jqa} and applies the instrument response from~\cite{Marsat:2018oam} to produce waveforms as they would appear in the A, E and T LISA TDI-1.5 data streams.

To make $d_{\rm obs}$ a function of also the time before merger, we must simulate observing the signal before it merges. To do this, we choose to terminate the integration $(h_{X}|h_{X})$
at some frequency corresponding to the frequency of the dominant gravitational mode a certain time before merger. This frequency-time correspondence can be computed
from the post-Newtonian expansion as shown in~\cite{Buonanno:2009zt}.

We then apply this to generate Figure~\ref{fig:motivation}, which shows
the time before merger at which we would be able to observe a \ac{MBHB}
signal as a function of chirp mass and source distance. This is plotted for the 3 possible sensitivity curves that we discuss in the previous subsection.
We observe that while the sensitive distance to premerger \ac{MBHB} signals is significantly reduced with respect to full bandwidth signals, there is a significant region of parameter space where detection is possible, even up to 14 days before merger.
For instance, a system with total mass of $10^6\,M_{\odot}$ might be observed up to $\sim 10\,$Gpc, 14 days before merger. 
We also note that, without any sensitivity below $10^{-4}\,$Hz our ability to observe premerger signals rapidly degrades for systems where the dominant (2,2) gravitational-wave emission mode has not yet evolved to $10^{-4}\,$Hz. This highlights the importance of low-frequency sensitivity for premerger observation of \ac{MBHB} signals.

\section{A method for identifying MBHB signals pre-merger}
\label{sec:search_method}

In this section, we lay out our method for identifying \ac{MBHB} signals
in advance of the merger, before demonstrating its effectiveness in
the following sections. Given that premerger signals are likely to have low signal to noise ratio we consider a matched-filter search approach, using
a suitable set of waveform filters and considering only the dominant (2,$\pm$2) gravitational-wave mode.
We begin this section by discussing how we construct our set
of waveform filters or ``template bank''.
We then discuss some complications in the
standard matched-filtering picture that arise when trying to identify
signals that exist at the very end of a given data set, and how we solve this using techniques developed within the GstLAL framework for low-latency identification of signals with ground-based observatories~\cite{Messick:2016aqy,Tsukada:2017cuf,Cannon:2020qnf}. We then discuss how we apply our set of waveform filters to observe premerger signals with minimal latency.

\subsection{Building a template bank of early-warning waveforms}

\begin{table}[t]
\begin{centering}
\begin{tabular}{|c| c|c|}
\hline
  Parameter & Limits & Distribution\\
  \hline
  Total Mass, $M_\odot$ & 1e6 -- 2e7 & uniform \\
  Mass ratio, $q$ & 1 -- 4 & uniform \\
  Component Spins, $\chi_1, \chi_2$ & -0.9 -- 0.9 & uniform\\
  Ecliptic Longitude, $\lambda$ & 0 -- $2\pi$ & uniform \\
  Ecliptic Latitude, $\beta$ & $-\pi / 2$ -- $\pi / 2$ & uniform in $\sin(\beta)$\\
  Inclination, $\iota$ & 0 -- $\pi / 2$ & uniform in $\cos(\iota)$ \\
  Polarisation & 0 -- $2 \pi$ & uniform \\  
\hline
\end{tabular}
\caption{Parameter space used for setting up the template bank for the early warning search
\label{tab:parameter_space}
}
\end{centering}
\end{table}

\begin{table}[t]
\begin{centering}
\begin{tabular}{|c|c|c|c|}
\hline
Time before merger & \multicolumn{3}{c|}{Number of templates} \\
\cline{2-4}
(days) & Optimistic & Pessimistic & Hard Cut \\
\hline
0.5 & 6,416 & 5,491 & 11,423 \\
1 & 7,997 & 5,755 & 13,713 \\
4 & 7,923 & 7,406 & 13,813 \\
7 & 5,580 & 5,589 & 9,738 \\
14 & 3,262 & 3,415 & 4,853 \\
\hline
\end{tabular}
\caption{Template bank sizes used in our search. A minimal match of 0.97 was
used to construct all of these banks.
\label{tab:template_bank_size}
}
\end{centering}
\end{table}

We first consider how many filter waveforms are needed to observe
premerger signals and how to construct our template banks. To create pre-merger waveforms when building the template banks, we generate a signal with frequency only extending up to the point corresponding to the desired time before merger when computing the overlap. We start signals from the frequency corresponding to 30 days before merger. 

To create a template bank to cover this parameter
space we use the technique of ``stochastic template placement''~\cite{Babak:2008rb, Harry:2009ea}. 
The stochastic placement relies on proposing a large number of points from within the parameter space according to a specified prior.
A point is accepted into the template bank if its overlap with all other waveforms already in the template bank is less than some specified threshold, in most cases chosen to be 0.97~\cite{Owen:1998dk, Babak:2006ty}.  
The algorithm continues to propose points until some predefined stopping condition is met, in this case we stop proposing points
when when we are accepting less than 5\% of new proposals.
Following~\cite{Allen:2005fk}, we maximize the overlap over the distance to the source and the orbital phase, and numerically search over the coalescence time.
The component masses, component spins,
sky location, polarization and inclination of the binary are then left as `intrinsic parameters' that we sample over when creating the template bank. For simplicity
we will consider only aligned-spin systems in this work, so that the
component spins are described by the two spin magnitudes. This leaves
us with 8 parameters to consider. In particular, we use the stochastic template bank algorithm available in the \texttt{PyCBC} software package~\cite{Usman:2015kfa, Kacanja:2024pjh}.

As in the previous section, the overlap is computed by generating waveforms using the \texttt{BBHx} package~\cite{Katz:2020hku, Katz:2021uax, michael_katz_2023_7791640}, which uses waveforms from the PhenomD waveform family~\cite{Khan:2015jqa} in the A, E and T LISA TDI data streams. We then compute the overlap of the waveform in \emph{both} A and E data streams~\footnote{We ignore the T data stream as it contains a negligible fraction of the signal power} with the A and E streams from all other waveforms already in the template bank. A single overlap is then computed from a weighted sum of the overlap in A and E according to
\begin{equation}
O = \frac{(O_A^2  \sigma_A^2 + O_E^2 \sigma_E^2)^{0.5}}{(\sigma_A^2 + \sigma_E^2)^{0.5}},
\end{equation}
where $\sigma_X$ is the optimal signal-to-noise ratio of the proposal point in the corresponding data stream and $O_X$ the overlap in A, or E, and $O$ the weighted sum overlap. If this weighted overlap is smaller than the overlap threshold for all waveforms currently in the bank then the proposal point is accepted into the template bank.

We then create template banks for signals 0.5, 1, 4, 7 and 14 days before merger. For all of these an overlap threshold of 0.97 was used.
The parameter space we consider is provided in Table~\ref{tab:parameter_space} and the resulting template bank sizes are given in Table~\ref{tab:template_bank_size}. We note that we have chosen a
somewhat restricted range of masses here. For systems with higher masses,
premerger detection becomes increasingly unlikely, and at lower masses
the signal is in band for much longer, reducing the need for rapid
identification. Nevertheless, the methods we present here remain valid
if one were to search in a broader range of masses.

There are some interesting features in the relative size
of the template banks. We note that the `hard cut' PSD consistently produces larger template banks than the other two, and we find the largest template banks are the ones 4 days before merger, with smaller banks required for both larger time before before, and shorter time.
For this work, the most important thing to note from these bank sizes is that they are small; The computational cost of filtering $\sim10^4$ templates is negligible. Nevertheless, in terms of the relative size of the `hard cut' PSD, we note that the sharp cutoff at $10^{-4}\,{\rm Hz}$, coupled with a termination frequency that in some parts of the parameter space is only slightly above this, results in a large number of waveforms in these parts of parameter space.
In short, we think this is a feature of the unphysical sharp transition we introduce into this PSD. The peak in template bank size at 4 days before merger seems to be a genuine feature. At the limit where one considers the full bandwidth signal almost all the power is contained in the last hour before merger, which can be covered to a 0.97 match with a relatively small number of templates. One doesn't have to use a template that correctly models the evolution over the full signal to be able to detect full bandwidth signals. Similarly, in the limit where we generate template banks a long time from merger the signal becomes short, and with little frequency evolution, requiring fewer templates. 

\subsection{A zero-latency whitening filter}

In Eq.~\ref{eq:mf_standard} we stated the standard matched-filtering
equation. This is often computed as a function of the signal
coalescence time by adding a $e^{-i 2 \pi f t}$ factor to the integral
and (if necessary) using an inverse Fourier transform to obtain the matched filter as a function of time. The same basic process can be applied when searching for \ac{MBHB} signals before merger. However, particular care must be taken to be able to analyse premerger signals in low-latency where the majority of the power is accumulating at the very end of the data being analysed.

\begin{figure}
    \includegraphics{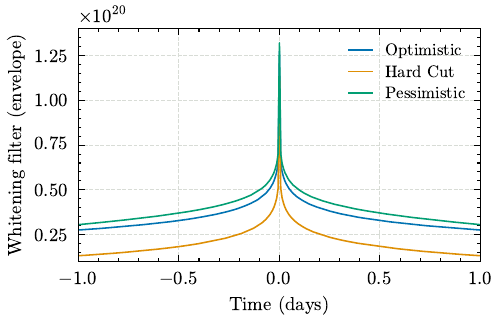}
    \includegraphics{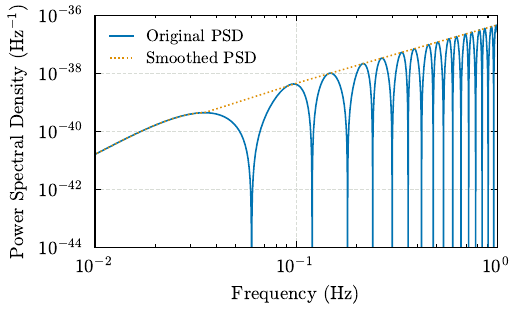}
    \includegraphics{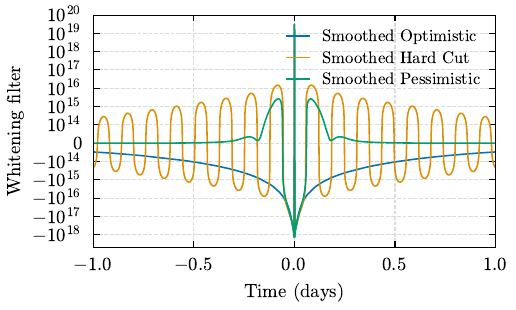}

    \caption{Top panel: Time-domain whitening filter of the LISA A/E sensitivity curve. Plotted as an envelope due to this containing a lot of short-duration structure, which is difficult to visualize. Middle panel: The high-frequency features of the LISA A/E \ac{PSD}, and the smoothing we use to remove these from the time-domain whitening filter. Bottom panel: Time domain whitening filter of the LISA A/E sensitivity curve after smoothing out the high-frequency features. Shown for both ``pessimistic" and ``optimstic" \acp{PSD}.}
  \label{fig:normal_psd_autocorrelation}
\end{figure}

Whitening a waveform, or data, in the frequency domain is performed by dividing by the square root of the \ac{PSD} (the \ac{ASD}). This is equivalent in the time-domain to \emph{convolving} with the inverse Fourier transform of the inverse of the \ac{ASD} ($\mathrm{IFFT}(S_N(f)^{-1/2})(t)$), also referred to as the whitening filter.
The ``length'' of the inverse \ac{ASD} in the time domain---the length of the whitening filter---is an important consideration.
If one has a whitening filter with a day duration, one can only filter data after waiting a day to allow sufficient data to accrue to correctly whiten the data. By this time the signal may well have already merged!

In the top panel of Figure~\ref{fig:normal_psd_autocorrelation} we show the time-domain whitening filter of the LISA A/E \acp{PSD} shown previously in Figure~\ref{fig:psd}.
There is a lot of structure in this whitening filter, which is difficult to visualize, and therefore we just plot the envelope of the amplitude of the whitening filter to illustrate that it extends over a day in each direction\footnote{One can access the full whitening filters in our data release, if interested.}.

This extended duration is primarily due to the high-frequency features, where the A/E TDI \ac{PSD} repeatedly drops to zero.
These features are not indicating that the sensitivity is approaching infinity at these frequencies, but correspond to points where noise will completely cancel during the TDI process, as seen in Figure~\ref{fig:psd}. 
The response of any gravitational-wave signals will also go to 0 at these frequencies, and there is no sensitivity at these points~\cite{Babak:2021mhe}.
As we are not searching for signals at the frequencies where these occur, we can remove these features from the whitening filter by smoothing the \ac{PSD}\footnote{One could also low pass filter the data, but we choose to smooth the PSD to avoid having to construct a zero-latency low-pass filter}.
In Figure~\ref{fig:normal_psd_autocorrelation} we also show
the whitening filter after removing the high-frequency features,
which has a much shorter characteristic length.
Nevertheless, this analytical \ac{PSD} is much smoother than what we expect in reality (see, for example, \acp{PSD} from ground-based instruments).
Therefore, it is fair to assume that the whitening filter length would be up to a day long for real LISA data analysis.

\begin{figure}
    \includegraphics{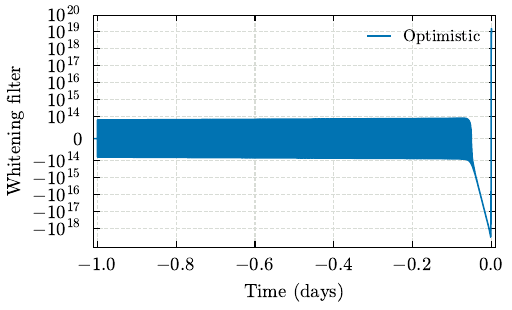}
    \includegraphics{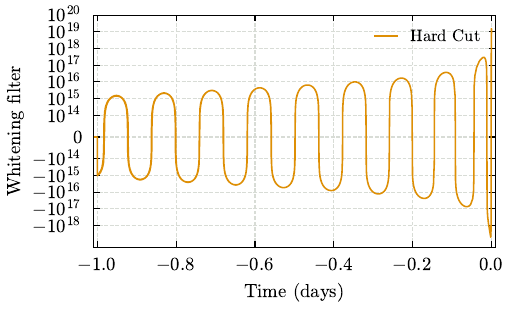}
    \includegraphics{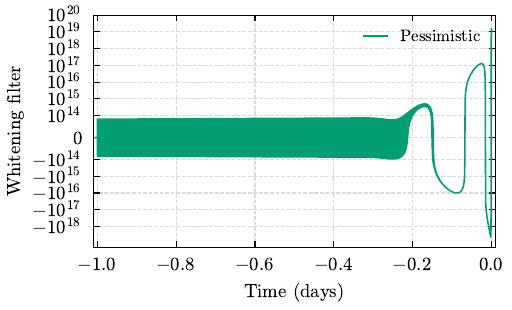}

    \caption{The zero-latency whitening filters used for the 3 \acp{PSD} shown in Figure~\ref{fig:psdzz}.}
  \label{fig:zero_latency_filters}
\end{figure}

A similar problem exists in low-latency searches with ground-based observatories. Indeed initial low-latency searches using the \texttt{GstLAL} search algorithm~\cite{Messick:2016aqy,Cannon:2020qnf} were limited to latency of around 20 seconds~\cite{Tsukada:2017cuf}, primarily because the code had to wait for 16 seconds of data to arrive before it could convolve the data with the \ac{PSD}\footnote{\texttt{GstLAL} performs its filtering directly in the time domain.}. To speed this up, methods developed in~\cite{ZLWF_Paper} were applied to create an approximate ``zero-latency whitening filter''~\cite{Tsukada:2017cuf}. This filter, shown in Figure~\ref{fig:zero_latency_filters} in the context of LISA, would be one-sided in the time-domain, so only data in the past is needed to filter data now. This method greatly reduced the latency of the \texttt{GstLAL} algorithm and allows identification of compact binary mergers in less than 10 seconds in ongoing LVK analyses~\cite{Tsukada:2017cuf,Cannon:2020qnf}.

We quickly summarize how the zero-latency filter is created, but point the reader to~\cite{ZLWF_Paper} for the full description, motivation and derivation. The steps to
compute the zero-latency filter are as follows
\begin{itemize}
    \item Compute the ``cepstrum'' time series of the PSD~\cite{RANDALL20173}; in our case, this is the inverse Fourier transform of the elementwise logarithm of the PSD.
    \item If $N$ is the length of the cepstrum, set the 0th and $N/2$ element to 0, and multiply all elements from the $N/2+1$ element to the $N-1$ element by $-1$.
    \item Take the Fourier transform of the resulting time series, to create the frequency series $\theta$.
    \item Multiply the PSD by the frequency series given by $e^\theta$.
    \item Take the inverse Fourier transform of this to obtain the zero-latency filter.
    \item Truncate the length of the zero-latency filter; in our case we allow a duration of one day.
\end{itemize}

\begin{figure}
    \includegraphics{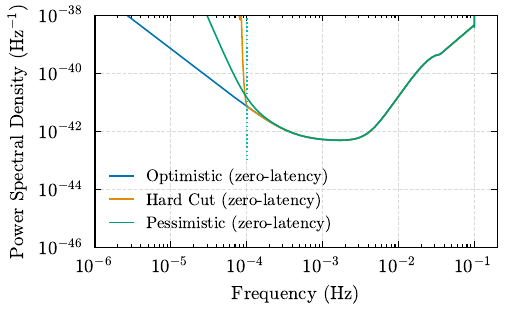}
    \caption{The effective \acp{PSD} of the zero-latency whitening filters. These are produced by Fourier transforming the whitening filters, taking the absolute value, inverting, and squaring to reproduce the \ac{PSD} response.}
  \label{fig:psdzz}
\end{figure}

We now use this method to construct zero-latency filters for the 3 sensitivity curves considered in this work.
Even though there is a truncation step in the method, it
is important that this is not removing much power from the
zero-latency whitening filter, or we will not accurately
whiten the data.
To achieve this, as already discussed, we smooth over the high frequency features, as shown in Figure~\ref{fig:normal_psd_autocorrelation}.
This is also why we chose to smooth out the abrupt termination of the ``hard cut'' PSD by multiplying the log of the \ac{PSD} with a sigmoid function to bring it to an arbitrary ``large" value ($10^{-36}$ $\mathrm{Hz}^{-1}$). This is shown in Figure~\ref{fig:psd}.
Having suitably conditioned our \ac{PSD}, we then directly apply the algorithm described using code available in the \texttt{GstLAL} software package~\cite{Cannon:2020qnf}.
This results in the whitening filters shown in Figure~\ref{fig:zero_latency_filters}. We can check that these have the intended frequency response by Fourier transforming the filters back to the frequency domain, and plotting the absolute value (squared).
This is shown in Figure~\ref{fig:psdzz}, which can be seen to accurately reproduce Figure~\ref{fig:psd}.

\subsection{Application of the zero-latency whitening filter}

We are then able to apply the zero-latency whitening filter 
to matched filter two vectors.
Consider that we have two time series $h(t)$ and $s(t)$. We assume that these terminate at some immediate point in time, either because the data has not yet been taken or because we are searching for a signal that has not yet reached merger. The matched filter is then performed according to the following steps
\begin{itemize}
    \item Convolve both time series with the zero-latency whitening filter to whiten both time series. For convenience, we perform this operation in the frequency domain, before applying an inverse Fourier transform to obtain the two whitened time series. 
    \item As with the ``standard'' whitening filter, the assumption that the data is periodic in the frequency domain will result in some data being invalid. We set to zero all data in the first day (corresponding to the filter length) of the whitened timeseries. Additionally, if the cutoff is not at the end of the dataset one must again set all data after the original cutoff time to zero. 
    \item The resulting timeseries can then be multiplied elementwise and summed to obtain the matched filter. One can quickly obtain the matched filter as a function of time-shift in the standard way, performing the multiplication in the frequency domain, adding an $e^{2 i \pi f t}$ factor, and an inverse Fourier transform.
\end{itemize}

We remind that we release and demonstrate the code we used in this work in our data release for the interested reader.

\section{Demonstrating the ability to find compact binary mergers before merger with LISA}
\label{sec:search_results}

\begin{table*}[t]
\begin{centering}
\footnotesize
\begin{tabular}{|c | cc cc c cc ccc ccc|}
\hline
\multirow{2}{*}{Signal Number} & $M_1$ & $M_2$ & $\chi_1$ & $\chi_2$ & $\lambda$ & $\beta$ & $\iota$ & Distance & $z$ & polarisation & phase & Time Offset & Full-band SNR \\
 & $10^6 M_\odot$ & $10^6 M_\odot$ &  &  & rad & rad & rad & Gpc & & rad & rad & seconds & \\
\hline
0 & 1 & 1 & 0 & 0 & 3.45 & 0.44 & 0.924 & 27.7 & 3.2 & 3.42 & 2.66 & 1050 & 1975.3 \\
1 & 2 & 0.5 & 0 & 0 & 1.51 & 1.07 & 1.52 & 17.4 & 2.1 & 1.87 & 2.32 & 3400 & 2125.0 \\
2 & 1 & 0.7 & 0.4 & -0.3 & 1.46 & -0.17 & 1.53 & 12 & 1.6 & 1.37 & 0.80 & 100 & 2049.8 \\
3 & 2.5 & 2.5 & 0.8 & 0.9 & 2.44 & -0.80 & 0.823 & 103 & 9.7 & 2.04 & 0.08 & 3100 & 1990.5 \\
4 & 10 & 10 & 0 & 0 & 1.59 & 0.15 & 1.38 & 11.3 & 1.5 & 4.67 & 1.02 & 1700 & 2396.2 \\
\hline
\end{tabular}
\caption{Detector-frame parameters used for our simulated signals. The time offset given is the gap between the actual merger time of the signal and the end of the dataset. This is introduced to allow for the fact that we will not know exactly when a signal will merge, and need to determine this (see wording in section~\ref{sec:search_results} for more detail). The injections' distances were all chosen to have a full-bandwidth signal-to-noise ratio of approximately 2000. The redshift given is a translation of the luminosity distance, and assumes a Plank15 cosmology. Here $M_1$ and $M_2$ are the detector frame masses, $\chi_1$ and $\chi_2$ are the component spins, and $\lambda$ and $\beta$ are the right ascension and declination respectively.
\label{tab:injections}
}
\end{centering}
\end{table*}

\begin{figure}
  \includegraphics{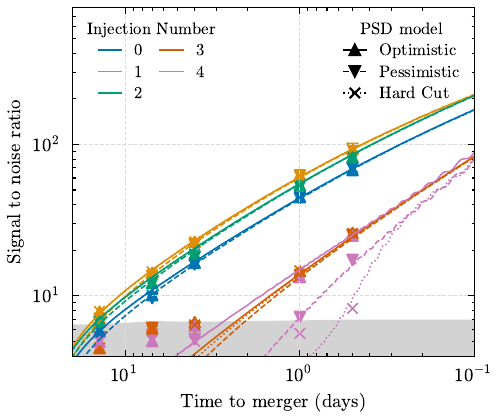}
  \caption{
  Simulated signals' optimal (lines) and recovered (markers) signal-to-noise ratio as a function of time to merger.
  The markers show results as in Table~\ref{tab:recovery} from our template bank search on the data.
  We also indicate with the grey line the approximate signal-to-noise ratio for a one-per-hundred-days \ac{FAR} given the mapping discussed around Figure~\ref{fig:snrifar}. Markers under this line have a good
  likelihood of being due to noise fluctuations, and not the real signal, and so we see a divergence between optimal and recovered signal-to-noise ratio under this line.
  }
  \label{fig:inj_opt_snr}
\end{figure}

To demonstrate our search technique, we consider 5 example signals that would
potentially be observable premerger. The
parameters of these are given in Table~\ref{tab:injections}. The example signals were chosen to demonstrate our method across the parameter space we consider, rather than being representative of any expected signal population. For example, Signal 3 was chosen to have significant spins, and Signal 4 was chosen to be at the most massive part of our parameter space.
The distances were chosen such that all 5 signals would have approximately the same optimal signal-to-noise ratio -- 2000 -- over the full bandwidth.
For each of these signals, we add the signal to simulated Gaussian noise and attempt to recover the signal 0.5, 1, 4, 7 and 14 days before merger.
Each of these signals was generated using the PhenomHM waveform including all available modes [(2,2), (2,1), (3,3), (3,2), (4,3) and (4,4)] and converted to the A and E TDI streams using the BBHx package~\cite{michael_katz_2023_7791640}. For the detection search, we remind that we are only using the (2,2) mode for recovery.

In Figure~\ref{fig:inj_opt_snr} we show the optimal signal-to-noise ratio of these signals in the (2,2) mode, as a function of time before merger.
As we can see, there is significant variation in the evolution of
signal-to-noise ratio of these signals before merger.
It is clear that signals 0, 1 and 2 would be observable at a greater
time before merger than signals 3 and 4 (dependent on the threshold for detection, which we will shortly demonstrate).
It is also important to highlight the importance of low frequency
sensitivity, in particular for signal 4, which may be observable
many days earlier if ``optimistic" sensitivity below
$< 10^{-4}$Hz is achieved.

We then perform a matched-filter search for these 5 signals using the zero-latency filter.
We begin by generating 30 days of Gaussian noise in the LISA A and E channels (using either the optimistic, pessimistic or hard cut PSDs).
We inject the relevant signal into this data stream. The merger of the
signal will always be in the last hour of the dataset.
The ``time offset" parameter in table~\ref{tab:injections} indicates
how much the merger is offset from the end of the dataset. We then
zero out the desired amount of data (noise and injected signal) at the end of the dataset. The process is then repeated for the template
waveforms in the template bank, except \emph{no} time offset is applied, so that the template waveforms are terminated exactly a certain amount of time before merger. The matched filter is then computed to obtain a signal-to-noise ratio time series.
We only consider the signal-to-noise ratio in a one-hour window, corresponding to the allowed time offset. This means that when we are searching for signals 4 days before merger, we are searching for signals that are between 4 days and 3 days, 23 hours before merger.
This is a somewhat arbitrary choice to demonstrate our method: We are analysing the data in small batches, and we chose to batch the data into hour-long stretches as a compromise between latency and computational cost. We note that GstLAL uses streaming to analyse data in realtime~\cite{Messick:2016aqy}, and while this could also be applied here, the non-streaming transmission of LISA data would likely not allow it in practice.

Having obtained signal-to-noise ratio time series in the A and E channels for a given template we then need to identify potential premerger signals. We implement a simplistic identification algorithm for this work: We look for the highest signal-to-noise ratio in one of the channels and then look in the other channel for the loudest signal-to-noise ratio within 100 seconds of the peak in the first.
We calculate the sum of squares for these, and then compare this to the same situation with the channels reversed; the larger of the two is recorded as the largest signal-to-noise ratio in that hour of data for that template.
When searching over our bank of templates, we simply take the largest signal-to-noise ratio over the full template bank and report this as the result of the search.
For the results shown here we only look for the signals in the hour of data where we know they are, but on real LISA data one would continually
repeat the process on each additional hour of data as soon as it was available.

\begin{table*}[t]
\begin{centering}
\footnotesize
\begin{tabular}{|c|c| c|c|c|c|c | c|c|c|c|c | c|c|c|c|c |}
\hline
\multirow{3}{*}{PSD} & & \multicolumn{5}{c|}{Optimal SNR}& \multicolumn{5}{c|}{Recovered SNR} & \multicolumn{5}{c|}{IFAR (years)} \\
& Signal & \multicolumn{5}{c|}{at T days premerger} & \multicolumn{5}{c|}{at T days premerger} & \multicolumn{5}{c|}{at T days premerger} \\
\cline{3-17}
& Number & $T=14$ & $T=7$ & $T=4$ & $T=1$ & $T=0.5$ & $T=14$ & $T=7$ & $T=4$ & $T=1$ & $T=0.5$ & $T=14$ & $T=7$ & $T=4$ & $T=1$ & $T=0.5$ \\
\hline
\multirow{5}{*}{Optimistic} & 0 & \ensuremath{5.87}& \ensuremath{11}& \ensuremath{16.8}& \ensuremath{44.4}& \ensuremath{69} & \ensuremath{5.84}& \ensuremath{10.1}& \ensuremath{16.6}& \ensuremath{44.9}& \ensuremath{67.7}& \ensuremath{0.0385} & $>100$ & $>100$ & $>100$ & $>100$ \\
 & 1 & \ensuremath{7.63}& \ensuremath{14.9}& \ensuremath{23.2}& \ensuremath{61}& \ensuremath{93.2} & \ensuremath{7.87}& \ensuremath{14.2}& \ensuremath{22.8}& \ensuremath{62.9}& \ensuremath{93}& \ensuremath{66.8} & $>100$ & $>100$ & $>100$ & $>100$ \\
 & 2 & \ensuremath{7.08}& \ensuremath{13.3}& \ensuremath{20.7}& \ensuremath{55.2}& \ensuremath{85.8} & \ensuremath{6.75}& \ensuremath{12.2}& \ensuremath{18.8}& \ensuremath{53.6}& \ensuremath{82}& \ensuremath{1.11} & $>100$ & $>100$ & $>100$ & $>100$ \\
 & 3 & \ensuremath{1.08}& \ensuremath{2.4}& \ensuremath{4.15}& \ensuremath{14}& \ensuremath{24.6} & \ensuremath{4.53}& \ensuremath{6.13}& \ensuremath{6.7}& \ensuremath{14.5}& \ensuremath{26}& \ensuremath{0.000314} & \ensuremath{0.0319} & \ensuremath{0.343} & $>100$ & $>100$ \\
 & 4 & \ensuremath{1.56}& \ensuremath{3.06}& \ensuremath{4.95}& \ensuremath{15.1}& \ensuremath{25.3} & \ensuremath{4.96}& \ensuremath{5.03}& \ensuremath{6.32}& \ensuremath{13.3}& \ensuremath{25}& \ensuremath{0.00154} & \ensuremath{0.000617} & \ensuremath{0.0786} & $>100$ & $>100$ \\
\hline
\multirow{5}{*}{Pessimistic} & 0 & \ensuremath{5.17}& \ensuremath{10.2}& \ensuremath{16.2}& \ensuremath{44}& \ensuremath{68.6} & \ensuremath{5.33}& \ensuremath{9.68}& \ensuremath{16.2}& \ensuremath{44.4}& \ensuremath{70.1}& \ensuremath{0.00669} & $>100$ & $>100$ & $>100$ & $>100$ \\
 & 1 & \ensuremath{6.8}& \ensuremath{14}& \ensuremath{22.4}& \ensuremath{60.3}& \ensuremath{92.5} & \ensuremath{7.08}& \ensuremath{13.6}& \ensuremath{22}& \ensuremath{62.4}& \ensuremath{93.4}& \ensuremath{5.38} & $>100$ & $>100$ & $>100$ & $>100$ \\
 & 2 & \ensuremath{6.51}& \ensuremath{12.7}& \ensuremath{20.1}& \ensuremath{54.9}& \ensuremath{85.4} & \ensuremath{6.68}& \ensuremath{12.2}& \ensuremath{18.1}& \ensuremath{53.2}& \ensuremath{81.8}& \ensuremath{1.19} & $>100$ & $>100$ & $>100$ & $>100$ \\
 & 3 & \ensuremath{0.587}& \ensuremath{1.68}& \ensuremath{3.36}& \ensuremath{13.3}& \ensuremath{24} & \ensuremath{6.16}& \ensuremath{6.14}& \ensuremath{6.23}& \ensuremath{13.7}& \ensuremath{25.3}& \ensuremath{0.163} & \ensuremath{0.0669} & \ensuremath{0.0601} & $>100$ & $>100$ \\
 & 4 & \ensuremath{0.163}& \ensuremath{0.481}& \ensuremath{1.1}& \ensuremath{7.12}& \ensuremath{16.1} & \ensuremath{5.26}& \ensuremath{5.56}& \ensuremath{5.18}& \ensuremath{7.28}& \ensuremath{17.3}& \ensuremath{0.00524} & \ensuremath{0.00638} & \ensuremath{0.001} & \ensuremath{2.21} & $>100$ \\
\hline
\multirow{5}{*}{Hard Cut} & 0 & \ensuremath{5.86}& \ensuremath{11}& \ensuremath{16.8}& \ensuremath{44.3}& \ensuremath{68.8} & \ensuremath{5.84}& \ensuremath{10.1}& \ensuremath{16.6}& \ensuremath{45.4}& \ensuremath{70.2}& \ensuremath{0.0372} & $>100$ & $>100$ & $>100$ & $>100$ \\
 & 1 & \ensuremath{7.63}& \ensuremath{14.8}& \ensuremath{23.2}& \ensuremath{60.8}& \ensuremath{92.8} & \ensuremath{7.9}& \ensuremath{14.4}& \ensuremath{22.8}& \ensuremath{62.9}& \ensuremath{94.6}& \ensuremath{46.8} & $>100$ & $>100$ & $>100$ & $>100$ \\
 & 2 & \ensuremath{7.06}& \ensuremath{13.3}& \ensuremath{20.6}& \ensuremath{55.2}& \ensuremath{85.5} & \ensuremath{6.8}& \ensuremath{12.4}& \ensuremath{18.9}& \ensuremath{54.1}& \ensuremath{82.3}& \ensuremath{1.05} & $>100$ & $>100$ & $>100$ & $>100$ \\
 & 3 & \ensuremath{0.111}& \ensuremath{1.76}& \ensuremath{3.79}& \ensuremath{13.8}& \ensuremath{24.3} & \ensuremath{4.55}& \ensuremath{5.89}& \ensuremath{6.45}& \ensuremath{14.6}& \ensuremath{25.8}& \ensuremath{0.000415} & \ensuremath{0.0243} & \ensuremath{0.189} & $>100$ & $>100$ \\
 & 4 & \ensuremath{1.58}& \ensuremath{1.75}& \ensuremath{1.94}& \ensuremath{3.53}& \ensuremath{9.67} & \ensuremath{4.96}& \ensuremath{5.08}& \ensuremath{5.02}& \ensuremath{5.67}& \ensuremath{8.27}& \ensuremath{0.00175} & \ensuremath{0.00109} & \ensuremath{0.000509} & \ensuremath{0.00472} & \ensuremath{79.6} \\
\hline
\end{tabular}
\end{centering}
\caption{The optimal and recovered signal-to-noise ratio (SNR) and inverse false-alarm rates (IFARs) for the 5 signals, 3 PSDs, and 5 times before merger that we consider.
\label{tab:recovery}
}
\end{table*}

\begin{figure}
  \includegraphics{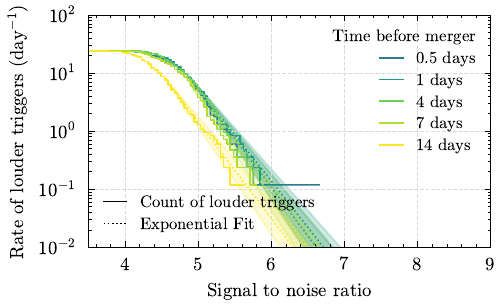}
  \caption{
      The mapping between signal-to-noise ratio and false alarm rate.
      Exponential fits (dotted lines) to the number of louder-ranked events in Gaussian noise (solid lines) are used to estimate the false alarm rate of injected signals above the background as reported in Table~\ref{tab:recovery}. 
  }
  \label{fig:snrifar}
\end{figure}

Finally, having identified the loudest signal-to-noise ratio in the hour window over the template bank, we must identify if it is a real premerger event or a noise fluctuation. We do this by computing a false-alarm rate, using a similar procedure to that described in~\cite{Davies:2022thw}. Specifically, we run
200 instances of the search in Gaussian noise, corresponding to 200 hours, and fit the distribution of the resulting signal-to-noise ratios to an exponential. This exponential is fit using the 83 loudest events (corresponding approximately to ten per day \ac{FAR}). One then has a mapping between the loudest signal-to-noise ratio in the hour window and a false-alarm rate. This mapping is calculated separately for each of our template banks---corresponding to different times before merger---and for each PSD, as the signal-to-noise ratio in noise alone will follow different distributions, largely dependent on the template bank size of Table~\ref{tab:template_bank_size}.
Figure~\ref{fig:snrifar} illustrates the background rates for the pessimistic 
search and the exponential fit we have used to map signal-to-noise to a false alarm rate.

Our results are presented in Table~\ref{tab:injections} and overplotted on Figure~\ref{fig:inj_opt_snr}.
Our search recovers signal-to-noise ratios for all the injections
that are close to the optimal signal-to-noise ratios. The optimal signal-to-noise does not include noise fluctuations, and so we always expect to see some variation from these optimal values.
Our search is capable of observing these premerger \ac{MBHB}
signals with high confidence as long as the signal has accumulated a signal-to-noise of roughly 8. 

We note that a number of our searches report a candidate that might
be considered ``marginal''; ones with recovered signal-to-noise ratios
of around 7. While such events would not be unambiguously identifiable as a premerger event, they would be significant enough to warrant rapid
follow-up. For example, having identified a potential event at 7 days premerger might warrant conducting a follow-up at 6 days before merger. A real event will quickly increase in significance, whereas a noise fluctuation will not.

\section{Parameter estimation for compact binary coalescences before merger in LISA}
\label{sec:PE}

\begin{table}
\begin{threeparttable}
\begin{centering}
\footnotesize
\begin{tabular}{|c | c |}
\hline
Parameter & Prior \\ \hline
$\mathcal{M}$ (signal 0) & $[8\times10^{5}, 9\times10^{5}]\;[\textrm{M}_{\odot}]$  \\
$\mathcal{M}$ (signal 1) & $[8\times10^{5}, 9\times10^{5}]\;[\textrm{M}_{\odot}]$  \\
$\mathcal{M}$ (signal 2) & $[7\times10^{5}, 8\times10^{5}]\;[\textrm{M}_{\odot}]$  \\
$\mathcal{M}$ (signal 3) & $[1\times10^{6}, 2.5\times10^{6}]\;[\textrm{M}_{\odot}]$  \\
$\mathcal{M}$ (signal 4) & $[8\times10^{6}, 10\times10^{6}]\;[\textrm{M}_{\odot}]$  \\
$q$ & $U[0.125, 1)$ \\
$\chi_1$ & $U[-0.99, 0.99)$ \\
$\chi_2$ & $U[-0.99, 0.99)$ \\
$\lambda$ & $U[0, 2 \pi)$ \\
$\beta$ & Cosine \\
$\iota$ & Sine \\
$d_\textrm{L}$ & Uniform in co-moving volume $[0.1, 200)$ [Gpc]\\
$\psi$ & $U[0, \pi)$ \\
$\phi_{\textrm{c}}$ & $U[0, 2 \pi)$ \\
$t_{\textrm{c}}$ & $U[-1, 1)$ [hours]\tnote{\textdagger}\\

\hline
\end{tabular}
\begin{tablenotes}
\item[\textdagger] For each injection, we randomly offset the prior from the injected value by up to $\pm 15$ minutes.
\end{tablenotes}
\caption{Priors used for performing inference in Section \ref{sec:PE}. The parameters are: chirp mass $\mathcal{M}$, mass ratio $q$, component spins $\chi_{\{1,2\}}$, the ecliptic longitude $\lambda$, ecliptic latitude $\beta$, inclination $\iota$, luminosity distance $d_{\textrm{L}}$, polarisation $\psi$, coalescence phase $\phi_\textrm{c}$ and time of coalescence $t_{\textrm{c}}$. $U[a, b)$ denotes a uniform prior over the semi-open interval $[a, b)$
\label{tab:priors}
}
\end{centering}
\end{threeparttable}
\end{table}

In this section, we describe how we can use the methodology described above to perform Bayesian parameter estimation for \ac{MBHB} systems detected before merger without latency and demonstrate our approach on the same example signals used in Section \ref{sec:search_results}.
We remind that other works have already shown how the measurement of parameters, in particular sky location, will improve as a system nears merger~\cite{Lops:2022ooj,Piro:2022zos}. In this section,
our primary motivation is to demonstrate that the method we have described \emph{can} be used in a real-world scenario to infer MBHB parameters without latency.
While we demonstrate the feasibility of the method, improvements and optimizations to the inference methodology we use will be needed before LISA launches.
In particular, we are not yet able to reliably infer parameters for systems less than a day out from merger, or to explore the interesting regime as the sky localization rapidly improves in the last day before merger.

\subsection{Configuration of parameter estimation runs}

We employ a Gaussian likelihood that is defined using the inner product from Eq.~\ref{eq:mf_standard} as
\begin{equation}
    p(s|\Theta, S_n(f))  \propto \exp \left\{ -\frac{1}{2}(s - h(\Theta)|s - h(\Theta))\right\},
\end{equation}
where $s$ is some noisy data, $h(\Theta)$ is a waveform with parameters $\Theta$ and $S_n(f)$ is the \ac{PSD}.
The only difference compared to the standard Gaussian likelihood used in e.g.~\cite{Thrane:2019}, is the use of the zero-latency whitening filter described in Section \ref{sec:search_method}.
This enables us to perform parameter estimation of \acp{MBHB} premerger without needing additional data for whitening.

We implement the likelihood in \texttt{PyCBC-inference}~\cite{Biwer:2018osg,Weaving:2023fji} and analyse the same five example signals described in Section \ref{sec:search_results} at times before merger where they have been found by the search with a \ac{FAR} of at least one-per-hundred-days.
For each signal, we analyse the LISA A and E channels, assuming coloured Gaussian noise distributed according to one of the \acp{PSD} described in Section \ref{sec:motivation}.
We employ \texttt{IMRPhenomHM}~\cite{London:2017bcn} as implemented in \texttt{BBHx}~\cite{michael_katz_2023_7791640} and perform inference using the nested sampling algorithm \texttt{nessai}~\cite{Williams:2021qyt,michael_j_williams_2024_10965503} which has been used for previous LISA analyses~\cite{Finch:2022prg,Buscicchio:2024asl}.

When performing inference, we place a two-hour prior on the time of coalescence.
We sample in the chirp mass and mass ratio and set the priors for chirp mass per injection. For all remaining parameters, we use the same priors for all injections. The full list of priors are detailed in Table \ref{tab:priors}.

\subsection{Parameter estimation results}

\begin{figure*}
    \centering
    \includegraphics[]{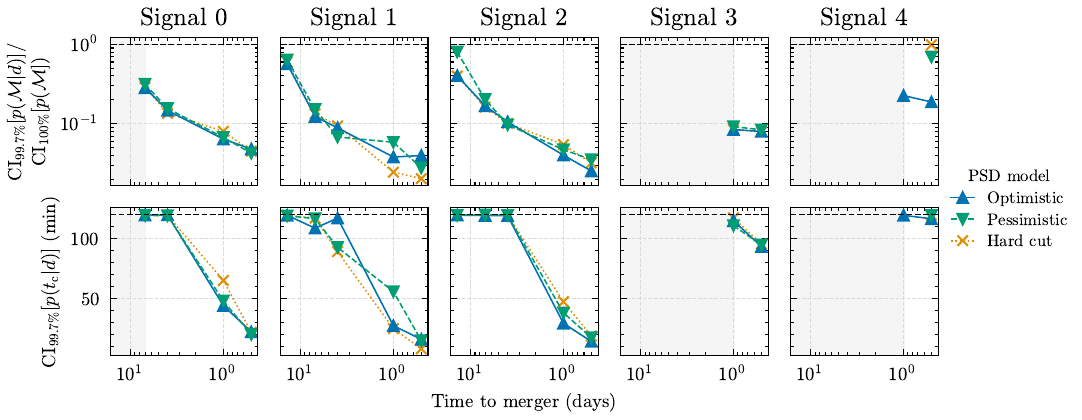}
    \caption{Width of the recovered posterior distributions for chirp mass (top) and time of coalescence (bottom) as a function of time to merger for each signal and \ac{PSD}. For chirp mass, results are shown as a fraction of the prior, computed as the $99.7\%$ confidence interval of the posterior $p(\mathcal{M}|d)$ divided by the width of the prior ($100\%$ confidence interval). For time of coalescence, the results show the  $99.7\%$ confidence interval of the posterior $p(t_c|d)$. The initial prior is indicated with a black dashed line. The shaded region denotes times at which the injection was not found by the search with any of the three \acp{PSD}, assuming a one-per-hundred-days \ac{FAR} threshold.
    \label{fig:pe:mass_time_post}
    }
\end{figure*}

\begin{figure}
    \centering
    \includegraphics{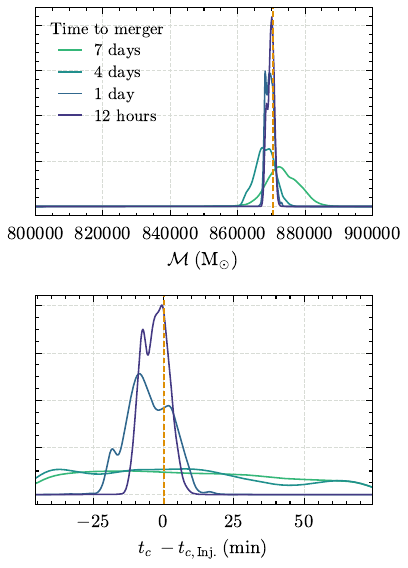}
    \caption{One--dimensional marginal posterior distributions for detector frame chirp mass $(\mathcal{M})$ and time of coalescence ($t_\text{c}$) for signal 0 as a function of time to merger when using the optimistic \ac{PSD}. The injected value is indicated with a dashed orange line. The posterior for time of coalescence has been shifted such that the injected value $t_\text{c,Inj}$ lies at zero.}
    \label{fig:pe:mass_tc_post_signal_1}
\end{figure}

Figure~\ref{fig:pe:mass_time_post} shows the evolution of the posterior distributions for detector frame chirp mass and time of coalescence as a function of time to merger for all combinations of signals and \acp{PSD}.
We find that the chirp mass is correctly inferred for all five signals irrespective of the choice of \ac{PSD}, however, our ability to constraint its value depends on the signal parameters.
For signals 0 to 3, the chirp mass is constrained to less than 50\% of the prior at 7 days before merger, irrespective of the choice of \ac{PSD}.
This is consistent with the results presented in Section \ref{sec:search_results}, where for all four signals, there are minimal differences in the recovered signal-to-noise ratio between PSDs.
For signal 4, which has a total mass of $2 \times 10^{7}\;\textrm{M}_{\odot}$, the chirp mass is only constrained when using the optimistic \ac{PSD} which we attribute to the additional signal-to-noise.
In Figure~\ref{fig:pe:mass_tc_post_signal_1} we show the one--dimensional marginal chirp mass posterior distribution for signal 0.
This demonstrates the iterative improvement in the inferred chirp mass as more of the signal is observed and also highlights the complex structure that arises from different degeneracies, which we discuss later.

We find the time of coalescence $t_\text{c}$ has a posterior distribution with comparable width to the prior (2 hours) at 7 days before merger for signals 0, 1 and 2.
However, at 1 day before merger the constraints improve significantly and the time of coalescence can be constrained to within less than an hour for the same signals.
At half a day to merger, the constraints improve by a factor of two and, for signal 1, $t_\text{c}$ is constrained to within 20 minutes.
For signal 3, $t_\text{c}$ is only constrained half a day before merger, and for signal 4 it is never constrained compared to the prior.

\begin{figure*}
    \centering
    \includegraphics{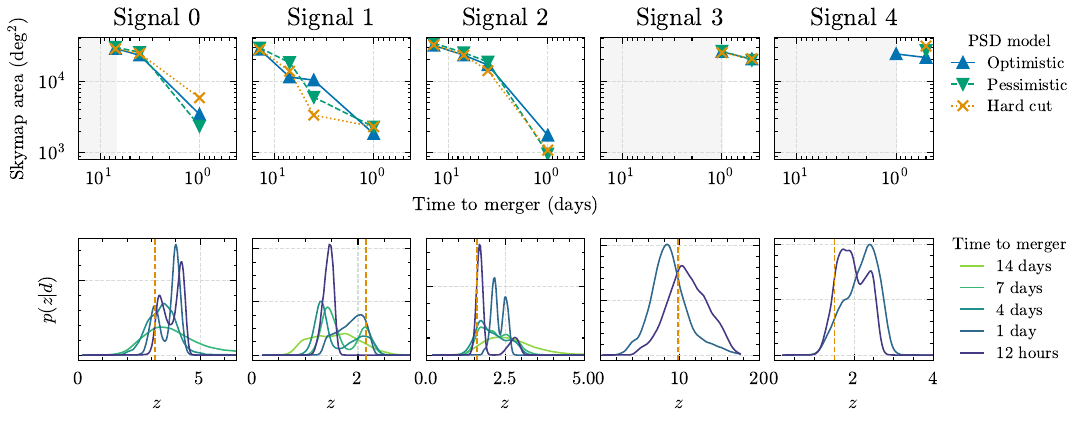}
    \caption{Inference results for the sky location and redshift. \textbf{Top:} area in square degrees of the 90\% contours of the sky maps for each signal as a function of the time before merger. Results are shown for each of the three \acp{PSD} described in Section \ref{sec:PE}. The shaded region denotes times at which the injection was not found by the search with any of the three \acp{PSD}, assuming a one-per-hundred-days \ac{FAR} threshold. Results at 0.5 days before merger for signals 0-2 are not included since we observe biases in these results, see Section~\ref{sec:PE:caveats}. \textbf{Bottom:} posterior distributions for redshift $p(z|d)$ for each signal as a function of time before merger when using the optimistic \ac{PSD}. The true value is indicated by the dashed orange line. The posterior distributions are only shown for signals that were found by the search assuming a one-per-hundred-days \ac{FAR} threshold. Results for the other \acp{PSD} as included in the data release.
    \label{fig:pe:sky_area}
    }
\end{figure*}

Figure~\ref{fig:pe:sky_area} shows our constraints on the sky localization of the source as a function of time to merger.
In most cases, the sky localization can be constrained to, at best, approximately 1000 deg$^{2}$ and we observe minimal differences between the different \acp{PSD}.
This is two orders of magnitudes larger than the field of view of electromagnetic facilities that may be online when LISA will be taking data~\cite{Colpi:2024xhw}.
This is consistent with previous studies e.g.~\cite{Lops:2022ooj,Piro:2022zos}, which found that for a \ac{MBHB} with total mass $3\times 10^{6} M_{\odot}$  at redshift 1, the sky localization at 1 day before merger was of order 1000 deg$^{2}$, however, these works also show that this localization will rapidly improve in the final day, improving to order 100 deg$^2$ one hour before merger before reaching smaller than 1 deg$^2$ after merger.

Figure~\ref{fig:pe:sky_area} also includes the inferred posterior distribution over redshift ($z$) for each signal when using the optimistic \ac{PSD}.
These show that the redshift can be constrained prior to merger, however, it is dependent on the signal-to-noise ratio of the signal and in most cases the uncertainties remain large even a day before merger.

\subsection{Caveats}\label{sec:PE:caveats}

\begin{figure*}
    \centering
    \begin{subfigure}[b]{0.24\textwidth}
        \includegraphics[width=1.1\textwidth]{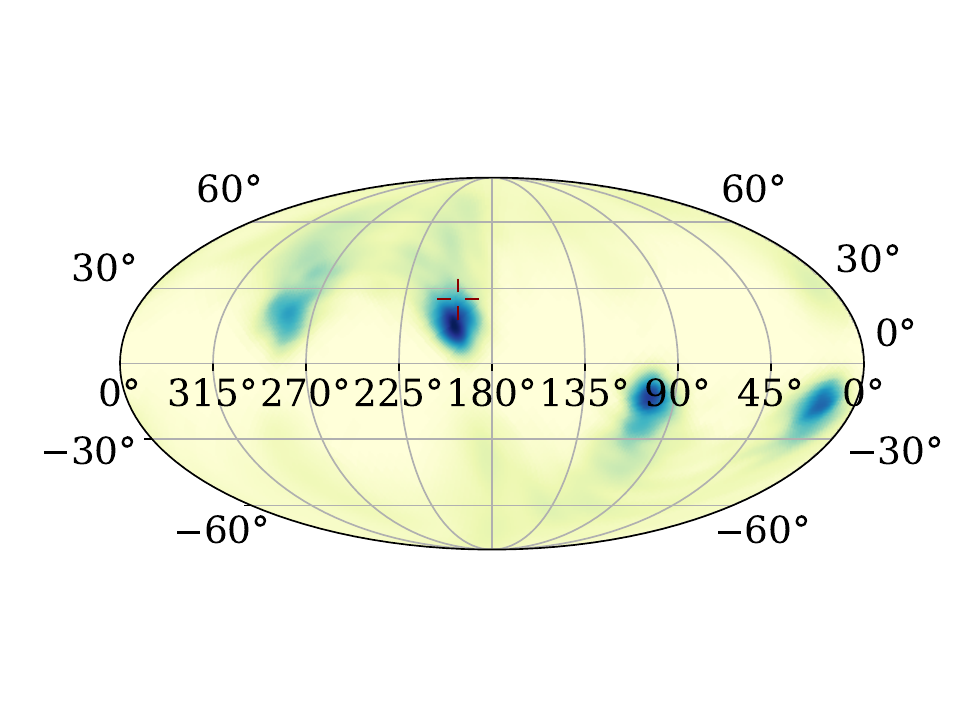}
        \caption{7 days}
        \label{fig:pe:example_skymap:a}
    \end{subfigure}
    \hfill
    \begin{subfigure}[b]{0.24\textwidth}
        \includegraphics[width=1.1\textwidth]{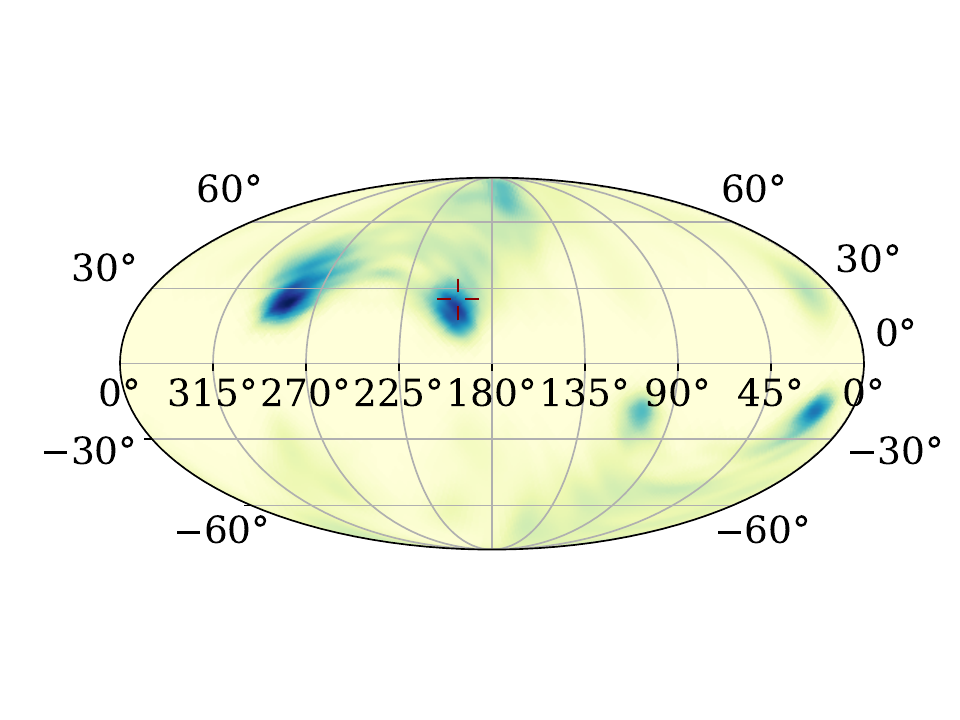}
        \caption{4 days}
        \label{fig:pe:example_skymap:b}
    \end{subfigure}
    \hfill
    \begin{subfigure}[b]{0.24\textwidth}
        \includegraphics[width=1.1\textwidth]{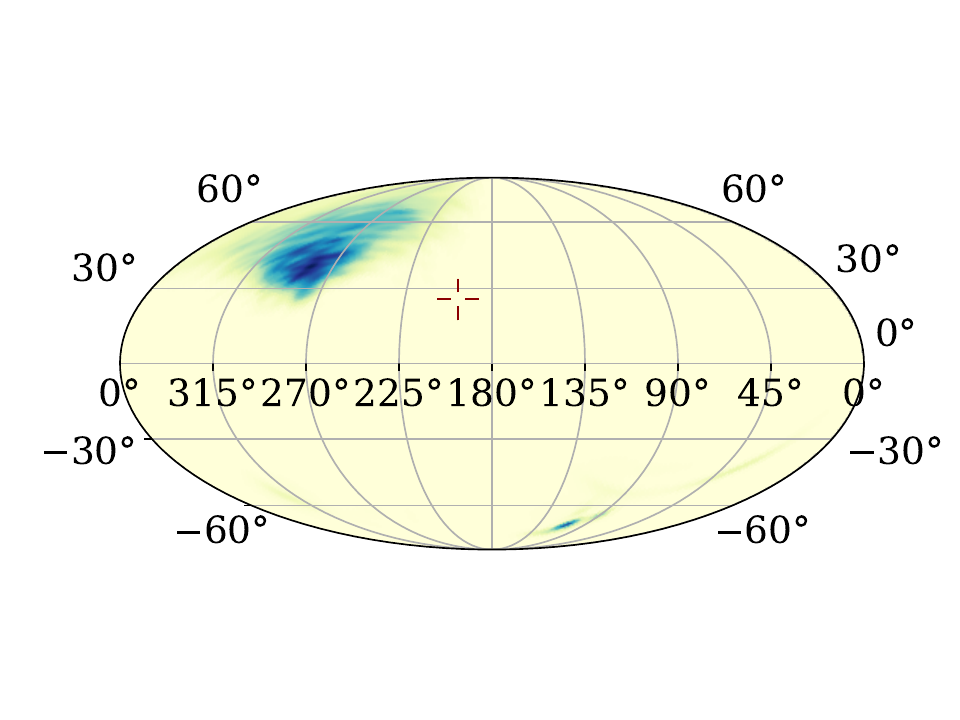}
        \caption{1 day}
        \label{fig:pe:example_skymap:c}
    \end{subfigure}
    \hfill
    \begin{subfigure}[b]{0.24\textwidth}
        \includegraphics[width=1.1\textwidth]{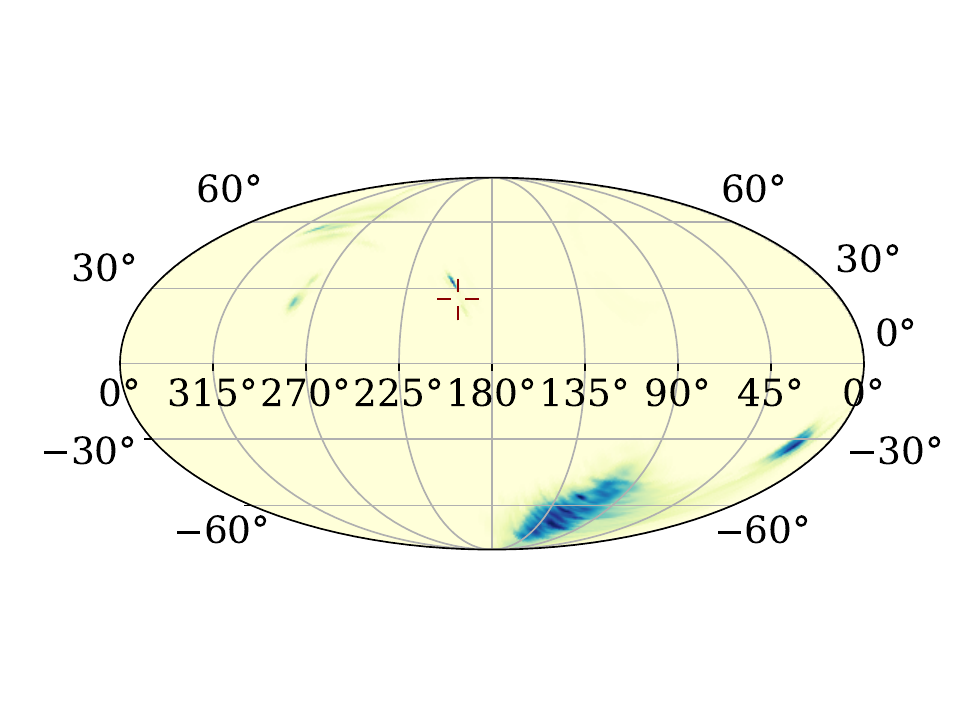}
        \caption{12 hours}
        \label{fig:pe:example_skymap:d}
    \end{subfigure}
    \caption{Inferred sky maps for signal 0 when using the hard cut \ac{PSD} at four times before merger. The true location is marked with a cross-hair. These results highlight the symmetries present in the posterior distributions, and how in some cases the posterior does not include support for the injected value but instead includes support for a mirror mode. See Section~\ref{sec:PE:caveats} for more discussion.}
    \label{fig:pe:example_skymap}
\end{figure*}

We find that sky localization parameters (ecliptic longitude and latitude) are not always inferred correctly when the signal-to-noise ratio becomes large.
We observe that there is a bias and signs of sampling issues for a subset of runs for signals 0-2 at 1 and 0.5 days before merger, when the signals have a signal-to-noise ratio of order 50 to 100.
In these cases, the posterior has support for a location on the sky that includes one of the mirror modes that arise due to the inherent symmetry in extrinsic parameter space described in \cite{Marsat:2020rtl} but does not have support for the injected value.
We show an example of this in Figure~\ref{fig:pe:example_skymap}.
We attribute this to sampling convergence issues that are a result of the multimodal space being sampled; in addition to the expected symmetries in the extrinsic parameter space, we also observe multimodality in the intrinsic parameter space which arises due to the lack of the late inspiral and merger phases in the data.
Given these biases, we do not include results for signals 0-2 at 0.5 days before in Figure~\ref{fig:pe:sky_area}, although an interested reader can find these in our data release.
This could be addressed, by incorporating the symmetries described in~\cite{Marsat:2020rtl} into the sampling as done in \cite{Weaving:2023fji}, or by modifying the sampling algorithm to be more robust when sampling multimodal problems, however, we leave this for future work.

These analyses take between one and five days to complete on 64 cores and scale with signal-to-noise ratio.
The wall time is dominated by the cost of evaluating the likelihood, which accounts for more than 95\% of the total wall time in all cases.
Therefore, in its current state, this implementation would not serve for issuing early warning alerts.
However, given likelihood evaluation is the dominant cost, reducing its cost, e.g. by leveraging graphical processing units, could significantly reduce the overall wall time.
Furthermore, following the first analysis, subsequent analyses do not make use of the results from previous analyses. Incorporating previous results into subsequent analyses could be another means of reducing the overall cost. We leave this for future work.

\section{Conclusion and Discussion}
\label{sec:conclusion}

In this work, we have demonstrated a novel end-to-end technique to 
observe \ac{MBHB} signals up to 14 days premerger in LISA data.
This technique uses a zero-latency whitening filter, which allows us
to include all the data that has been taken at a given point in time, and not lose data due to filter corruption from the whitening filter. We have shown that one can construct a template bank of signals
to find such premerger signals, and demonstrate that we can confidently
recover these premerger signals once they have accrued a signal-to-noise ratio in the LISA data of 8. The template bank sizes are of order 10,000 signals, and therefore represent a negligible computational
cost for a search that would not be applied on real data until the mid to late 2030s.

We also demonstrate the ability to use the zero-latency whitening filter
in Bayesian inference, and quantify the ability to localize the source on the sky, and determine the time to coalescence, in advance of the merger. We find that while we can constrain the merger time to 10s of minutes when over a day out of merger, and recover the chirp mass to around 1\% accuracy, the sky localization remains large (1000 square degrees and more) even half a day before merger.

While carrying out multi-messenger observations of a \ac{MBHB} premerger, and during the merger, will be challenging, the first, and most important step, is to provide as much notice that the event is going to happen. Thus notified, LISA could be put into a ``protected period" to ensure that LISA is taking data at the time the merger is going to happen~\cite{Colpi:2024xhw}. In addition, the sky localization will rapidly improve as the signal nears merger and this will continue to rapidly improve in the final 0.5 days before merger. Developing techniques to quickly improve upon sky localization in these final 12 hours will be crucial to maximizing the prospect of rapid electromagnetic observations.

The implementation of the technique used for this paper was sufficient to demonstrate the applicability of the method, and, given the small template bank sizes, is sufficient to allow the rapid detection of premerger signals. However, especially in the context of Bayesian inference, it would be greatly beneficial to optimize the computational cost of the likelihood, which is dominated by the generation of the whitened waveforms for filtering. Most obviously, the use of GPUs to speed up the likelihood computation would be highly beneficial.

A complication we encountered with this work was the use of frequency-domain waveforms and wraparound. As the Phenom waveforms are intrinsically periodic, we cannot ask the waveform to just stop at a certain point in time. We can generate the waveform, Fourier transform, and zero data out, but as the waveform wraps around, we often have (for example) ringdown appearing in the first day of the signal, which has wrapped around from the end. A fast, accurate time-domain waveform would help to alleviate these effects.

\begin{acknowledgments}
This work was supported by a series of UKSA grants supporting the UK's contribution to LISA's Ground Segment activities.
GCD, IH and MJW acknowledge support from ST/X002225/1 and ST/Y004876/1. LN and CH acknowledge support from the Future Leader's Fellowship scheme MR/T01881X/1.
J-BB and GW acknowledge support from ST/Y004906/1.
AK, HM, CM and AV acknowledge support from ST/Y004922/1, ST/V002813/1 and ST/X002071/1.
G.P. acknowledges support from a Royal Society University Research Fellowship URF{\textbackslash}R1{\textbackslash}221500 and RF{\textbackslash}ERE{\textbackslash}221015.

The authors are grateful for computational resources provided by Cardiff University, and funded by STFC awards supporting UK Involvement in the Operation of Advanced LIGO. Numerical computations were also carried out on the Sciama High Performance Computing (HPC) cluster, which is supported by the Institute of Cosmology and Gravitation, the South-East Physics Network (SEPNet) and the University of Portsmouth.

\textit{Software} This work made use of the following software packages: \texttt{BBHx}~\cite{michael_katz_2023_7791640}, \texttt{corner}~\cite{corner}, 
\texttt{GstLAL}~\cite{Cannon:2020qnf},
\texttt{ligo.skymap}~\cite{Singer:2015ema}, \texttt{matplotlib}~\cite{Hunter:2007}, \texttt{nessai}~\cite{Williams:2021qyt,michael_j_williams_2024_10965503}, \texttt{numpy}~\cite{harris2020array}, \texttt{pycbc-inference}~\cite{Biwer:2018osg}, \texttt{scipy}~\cite{2020SciPy-NMeth}, \texttt{seaborn}~\cite{Waskom2021}
\end{acknowledgments}

\bibliography{bibliography}

\end{document}